\documentstyle[12pt,jeep]{article}
\numberbysection
\date{}
\begin{document}
\title{\bf Nonautonomous Hamiltonians}
\author{
A. Soffer \thanks{Department of Mathematics, Rutgers University, New
Brunswick, NJ} \hspace{.05 in}
 and
  M.I. Weinstein \thanks{Department of Mathematics,
  University of Michigan, Ann Arbor, MI
   }}
   \baselineskip=18pt
   \maketitle
\def\M{\Gamma^{-\alpha}}
\def\un{\underline}
\def\nn{\nonumber}
\newcommand{\no}{\nonumber}
\newcommand{\be}{\begin{equation}}
\newcommand{\ee}{\end{equation}}
\newcommand{\ba}{\begin{eqnarray}}
\newcommand{\ea}{\end{eqnarray}}
\newcommand{\ve}{\varepsilon}
\newcommand{\nit}\noindent
\newcommand{\D}\partial
\def\Pc{{\bf P_c}}
\newcommand{\lan}\langle
\newcommand{\ran}\rangle
\newcommand{\wpl}{w_+}
\newcommand{\wmi}{w_-}
\newcommand{\la}{\lambda}
\newcommand{\ra}{\rightarrow}
\newcommand{\Z}{{\rm Z\kern-.35em Z}}
\newcommand{\bP}{{\rm I\kern-.15em P}}
\newcommand{\Q}{\kern.3em\rule{.07em}{.65em}\kern-.3em{\rm Q}}
\newcommand{\R}{{\rm I\kern-.15em R}}
\newcommand{\h}{{\rm I\kern-.15em H}}
\newcommand{\C}{\kern.3em\rule{.07em}{.55em}\kern-.3em{\rm C}}
\newcommand{\T}{{\rm T\kern-.35em T}}
\newtheorem{theo}{Theorem}[section]
\newtheorem{defin}{Definition}[section]
\newtheorem{prop}{Proposition}[section]
\newtheorem{lem}{Lemma}[section]
\newtheorem{cor}{Corollary}[section]
\newtheorem{rmk}{Remark}[section]
\begin{abstract}
\nit
We present a theory of resonances for a class of
non-autonomous Hamiltonians 
 to treat the structural instability   
 of spatially localized and time-periodic 
 solutions associated with an  
 unperturbed autonomous Hamiltonian.
  The mechanism of instability is radiative decay, due to 
 resonant coupling of the discrete modes to the continuum modes by the
 time-dependent  
  perturbation. This results in a slow transfer of energy from the discrete
 modes to the continuum. 
  The rate of decay of solutions is 
 slow and hence the decaying bound states can be viewed as metastable.
 The ideas are closely related to the authors' work on (i) a time dependent
 approach to the instability of eigenvalues embedded in the continuous
 spectra, and (ii) resonances, radiation damping  and instability in
 Hamiltonian nonlinear wave equations.
 The theory is applied to a general class of Schr\"odinger equations. 
   The phenomenon of ionization may be viewed as a resonance problem  
   of the type we consider 
 and we apply our theory to find the rate of ionization,
 spectral line shift 
 and local decay estimates for such Hamiltonians.
\end{abstract}
\thispagestyle{empty}
\vfil\eject

\section{Introduction}

There are many dynamical systems of interest for which the 
dynamics
can be viewed as given by a Schr\"odinger type equation generated by an
operator $H(t)=H_0+W(t)$. Here, $H_0$ is a time-independent (autonomous)
part, assumed to be 
 self-adjoint  on a complex 
  Hilbert space ${\cal H}$,  and $W(t)$ denotes
  time-dependent (non-autonomous)
  perturbation. We consider the initial value problem for the
  Schr\"odinger equation for $\phi : t\mapsto \phi(t)\in{\cal H}$:
\be
i\D_t\phi\ =\ \left(H_0\ +\ W(t)\right)\phi,
\label{eq:abstracttdse}\ee
with initial condition $\phi(0)=\phi_0$ in a suitable subspace of 
 ${\cal H}$.
An important special case, studied in section 3, is:
\be
i\D_t\phi(t,x)\ =\ \left(H_0\ +\ W(t,x)\right)\phi(t,x),
\label{eq:tdse}\ee
where $H_0=-\Delta + V(x)$, with $V(x)$ real-valued and 
 $W(t,x)$ is a smooth function which is spatially localized and
time-periodic or quasi-periodic.
Equations of this type play an important role in tunneling and ionization 
 physics,
 and
are relevant to an understanding of many nonlinear problems. This is 
discussed  later in this section.

In this paper, we consider the situation where the unperturbed equation,
 corresponding to the case $W\equiv0$,
has a time-periodic solution in ${\cal H}$,
 {\it e.g. a bound state}.
 We study the {\it structural instability} of the bound state
 due to a perturbation  $W(t)$, which is assumed to be periodic. 
 The analysis is carried out for the special case $W(t) = \cos(\mu t)\ \beta$, with
 $\beta^*=\beta$; see section 2. The case of more general perturbations $W(t)$, 
 which are periodic or quasi-periodic, as well as the case when the unperturbed  
 operator $H_0$ has more than bound state can be treated by our approach. 
 We indicate briefly in section 4 the modifications    
 required in the analysis. 

 Viewed in terms of the coordinates
  of the unperturbed dynamical system, 
  the perturbation $W(t)$ has the effect of coupling the
 bound state to the continuum modes. 
  The mechanism of instability is a
 slow transfer of energy from the bound state part of the solution to
 the continuum modes and an associated decay due to radiation of energy
 to infinity.
  We develop a theory which yields a detailed picture of
 the intermediate and long time behavior 
  of solutions to the initial value problem. Our main results are
  presented in Theorems 2.1 and 2.2. A consequence is that the 
   solution of the initial
  value problem for the perturbed dynamical system with initial data
  given by the time-periodic 
   bound state of the unperturbed problem, is characterized
  by transient exponential decay $\sim e^{-\Gamma |t|}$
   where $0<\Gamma = {\cal O}(|||W|||^2),$ for $|||W|||$ small,  and by
   algebraic decay as $|t|\to\infty$.
  We derive explicit formulae for the {\it lifetime}, 
   $\tau\equiv\Gamma^{-1}$ (eqn. (\ref{eq:tdfgr}))  and  
  the {\it spectral line shift} (eqn. (\ref{eq:rhoexpansion})).

The approach we take is closely related to our work on quantum
resonances \cite{kn:TDRT0}, \cite{kn:TDRT1} 
 and on the radiation damping of localized states for nonlinear
 wave equations \cite{kn:rdamping}. We now illustrate the main idea in
 the current context. Consider a linear Schr\"odinger equation:
 \be
 i\D_t\phi(t,x)\ =\ \left( -\Delta + V(x)\right)\phi(t,x)\ 
  +\ \cos(\mu t)\ \beta(x)\phi(t,x).\label{eq:model}
  \ee
We assume that $-\Delta + V(x)$ has exactly one eigenvalue at
$\lambda_0<0$, with corresponding eigenfunction $\psi_0(x)$. Therefore,
the unperturbed ($\beta\equiv0$) problem has a spatially localized and
time-periodic solution, $\psi_0(x)e^{-i\lambda_0 t}$. We assume that the
continuous spectrum of $H_0$ is equal to the real nonnegative
half-line. The perturbation $W(x,t) = \beta(x) \cos(\mu t)$ is
assumed  to be real-valued, to decay sufficiently rapidly as $|x|\to\infty$,
and to be small in an appropriate norm; see section 3.
\medskip

 \nit We now ask the following
 question: 
  {\it Does the bound state solution of the unperturbed problem  
 persist under small perturbations of this type?}

As indicated above, we find that generic perturbations lead to a slow
radiative decay of solutions. To illustrate the
mechanism of decay, we first observe that it is natural in  
the unperturbed case to decompose the solution as:
\be
\phi(t,x)\ =\ a(t)\psi_0\ +\ \phi_d(t,x),\ \
\left(\psi_0,\phi_d(t)\right)\ =\ 0.
\label{eq:model-decomp}
\ee
Then, we have that 
\ba
i\D_t a(t)\ &=&\ \lambda_0\ a(t),\nn
\\
 i\phi_d(t,x)\ &=&\ \left(-\Delta + V(x)\right)\  \phi_d(t,x),
\label{eq:model-decoupledeqns}
\ea
with initial conditions
\be a(0)\ =\ \left(\psi_0,\phi(0)\right),\ \ 
 \phi_d(0)\ =\ \Pc\phi(0)\ \equiv\  
 \phi(0)\ -\ \left(\psi_0,\phi(0)\right)\psi_0.\nn\ee
 $\Pc$ is the projection onto the continuous spectral part of
the operator $H_0$. The equations (\ref{eq:model-decoupledeqns})
decouple and we have that all solutions of  (\ref{eq:model-decomp}) are  
of the form (\ref{eq:model-decomp}) with:
\nit (i)\  $a(t)\ =\ e^{-i\lambda_0 t }\ a(0)$, and 
 under suitable hypotheses on $V(x)$,
\nit (ii) $\phi_d(t,x)$  decays (dispersively) as $|t|\to\infty$, in the
sense that $\|\phi_d(t,\cdot)\|_p\to0$,\ ($p>2$) or for some $\sigma>0$,  
$ \| \lan x\ran^{-\sigma}\phi_d(t,\cdot)\|_{L^2}\to0$.

How is the dynamics effected by a small spatially localized and
time-periodic perturbation ($\beta\ne0$)?
 Substitution of (\ref{eq:model-decomp}) into (\ref{eq:model}), 
  leads to a {\it weakly coupled} dynamical system for $a(t)$ and $\phi_d(t)$. 
   This system is derived and studied in sections 4-6.
  A simplification of the resulting system which we
  shall use to illustrate the main idea is the following:
\ba
i\D_ta(t)\ &=&\  \lambda_0\ a(t)\ +\ \left(\ \beta\psi_0\ ,\ \phi_d(t)\ \right)\
\cos(\mu t)\ ,\nn\\
i\D_t\phi_d(t)\ &=&\ -\Delta\phi_d(t)\ +\ a(t)\ \cos(\mu t)\ \beta \psi_0.
\label{eq:model-coupledsystem}
\ea

The system (\ref{eq:model-coupledsystem}) can viewed as an infinite
dimensional dynamical system consisting of two components: (i) a finite
dimensional part, described an "oscillatory", with 
 complex amplitude, $a(t)$,  coupled to 
(ii) an infinite dimensional dispersive wave equation on $\R^n$, governing
$\phi_d(t,x)$.
It is simple to check that the system (\ref{eq:model-coupledsystem})
 has the conserved energy:
\be
{\cal E}[a,\phi_d](t)\ \equiv\ |a(t)|^2\ +\ \int\ |\phi_d(t,x)|^2\
dx\ =\ {\cal E}[a(0),\phi_d(0)].\label{eq:model-energy}
\ee
The first term in (\ref{eq:model-energy}) corresponds to the oscillator
energy and the second to the wave energy. 

Intuitively, a spatially localized finite energy disturbance of the rest state 
 should result in motions induced in the oscillator as well
as waves propagating "outward" and eventually escaping any compact
set (a characteristic of dispersive waves). 
 Due to the coupling ($\beta\ne0$)
  of the oscillator and waves, we
 expect the 
energy of the oscillator, $|a(t)|^2$, to decay toward zero 
 as time progresses. This energy is
 transferred to the continuum modes, while the total energy of
the system, ${\cal E}[a,u]$, remains constant in time. This is what is proved
under suitable hypotheses in Theorems 2.1 and 2.2. 

The way in which we establish {\it energy transfer} from the discrete to the
continuum modes is by showing that one can transform 
a system of the type (\ref{eq:model-coupledsystem}) to a
 {\it normal form}, in which {\it
internal damping}, which drives the energy transfer, is made explicit. 
 This damping is due to a resonant
coupling of the discrete oscillator mode with the continuum modes. 
We now explain the origin of this effect.

Viewing $\beta$ as small, we set $a(t)\ =\ e^{-i\lambda_0 t}\ A(t)$,
 where $A(t)$ is "slowly varying", ($\D_tA(t)\ =\ {\cal O}(\beta)$). 
 In terms of the amplitude, $A(t)$, 
 the system (\ref{eq:model-coupledsystem})  becomes: 
\ba
\D_t\ A(t)\ &=&\ -i\ \left(\ \beta\psi_0\ ,\ \phi_d(t)\ \right)\ \cos{\mu t}\
e^{i\lambda_0 t},\label{eq:this1}\\
i\D_t\phi_d\ &=&\ -\Delta\ \phi_d\ +\ A(t)\ e^{-i\lambda_0 t}\ 
 \cos(\mu t) \beta\ \psi_0\label{eq:this2}
 \ea
 Since the variation of $A(t)$ is slow, 
  the source term in (\ref{eq:this2}) is essentially 
   quasi-periodic in time. One of its 
 frequencies is 
 $\lambda_0+\mu$, which we assume to be 
 strictly positive, {\it i.e.} $\lambda_0 +\mu\in\sigma_{cont}(-\Delta)$. 
  That is,  the non-autonomous parametric forcing of frequency
 $\mu$ induces a  resonant forcing of the  wave-field $\phi_d$.
The character of solutions to (\ref{eq:this1}-\ref{eq:this2}) can be deduced 
 by first solving
(\ref{eq:this2}) for $\phi_d$ and 
 then substituting the result into  (\ref{eq:this1}).
The result is a reduction to an {\it ordinary differential equation} for $A(t)$.
This is reminiscent of the center manifold approach, commonly used in dissipative
systems (see, for example, \cite{kn:Carr}, \cite{kn:VI}) 
 and recently applied in a Hamiltonian
context \cite{kn:pillet-wayne}.
 
 The asymptotic behavior of the ordinary differential equation governing $A(t)$
 can be deduced by a asymptotic evaluation of the right hand side of
 (\ref{eq:this1}), considering carefully the contribution to $\phi_d$  coming
 from a neighborhood of the resonant energy, $\lambda_0+\mu$. 
 \ba
&&-i\ \left(\ \beta\psi_0\ ,\ \phi_d(t)\ \right)\ \cos{\mu t}\ e^{i\lambda_0 t}
  \nn\\
 &=&\ -{i\over4}\ \left(\ \beta\psi_0\ ,\ \int_0^t\ e^{i\Delta (t-s)}
 e^{-i(\lambda_0+\mu)s} A(s)\ ds\ \beta\psi_0\ \right)\ e^{i(\lambda_0+\mu)t} 
 \ +\ {\rm nonresonant\ terms} 
  \nn\\
\ &=&\ -{1\over4}\ \left(\ \beta\psi_0\ ,\ (-\Delta -\lambda_0 -\mu -i0)^{-1}\
\beta\psi_0\ \right)\ A(t)\ +\ {\rm nonresonant\ terms}
  \label{eq:model-expand}
  \ea
  Substitution of this result into the equation
  for $A(t)$ yields:
  \be
  \D_tA(t)\ =\ \left( -\Gamma\ +\ i\Lambda\ +\ \rho(t)\ \right)\ A(t)\,\
   \ \ t>0\label{eq:model-damping}\ee
  where  
  \be
  \Gamma\ =\ \pi\ \left(\ \beta\psi_0\ ,\ \delta(-\Delta-\lambda_0-\mu)\
  \beta\psi_0\ \right).
  \label{eq:model-fgr}
  \ee
  $\Lambda$ is a real constant (the {\it Lamb shift}) and  ${\cal O}(\beta^2)$ 
   and $\rho(t)$ is  a bounded
  oscillatory function of $t$ which is ${\cal
  O}(\beta^2)$. 

The right hand side of (\ref{eq:model-fgr}) is defined 
 using the spectral theorem for
$-\Delta$ on $L^2(\R^n)$. In terms of the Fourier transform, ${\cal  F}[z](\xi)\
=\ \int\ e^{-i\xi\cdot x}\ z(x)\ dx$, we find:
\be
\Gamma\ =\ {1\over16}\ (\lambda_0+\mu )^{-{1\over2}}\ \left( \left|{\cal
F}[\beta\psi_0]\left(\sqrt{\lambda_0+\mu}\right)\right|^2\ +\  \left| {\cal
F}[\beta\psi_0]\left(-\sqrt{\lambda_0+\mu}\right)\right|^2 \right).
\label{eq:model-fgr-ft}
\ee
   The constant 
   $\Gamma$ is always non-negative. If it is strictly positive, 
  then (\ref{eq:model-damping}) is  a {\it damped oscillator}, with
  damping coefficient $\Gamma$ resulting from
   coupling of the discrete and continuum modes.
 In our general theorems, 
    Theorems 2.1 and 2.2,  we have hypothesis {\bf (H6)}, that the
	analogous general expression for $\Gamma$ is strictly positive.
	This holds generically in the space of perturbations $W$ under
	consideration.

\nit{\it Emergence of irreversible behavior in a time-reversible system}

Thus, the original dynamical system, (\ref{eq:model}), which is
Hamiltonian and, in particular,  time-reversible ( invariant under the
transformation $\phi(t,x)\mapsto\overline\phi(-t,x)$ ), is equivalent to a
dynamical system with some {\it apparent} time-asymmetry.
\footnote{  
 This is related to a "paradox" raised in the late nineteenth
century by Zermelo and others concerning the irreversibility of
macroscopically observed behavior in physical systems and the
reversibility of the microscopic equations of motion.
 This was addressed in the context of
classical systems via a  statistical approach in the work of Maxwell and
Thomson and  culminated in the work of Boltzmann. In quantum systems, for
example in the quantum measurement theory, extensions of these ideas
apply. The source of the emergence of irreversibility, is the coupling of
systems with few degrees of freedom to systems with many. 
 For a discussion, see \cite{kn:Lebowitz}.}     
This is related to the {\it $\varepsilon$ - prescription} discussed in
section 4; see, in particular, Proposition 4.2. At the heart of this
apparent paradox is that
 the energy of the oscillator, which is coupled to the
continuum, is propagated out to infinity
 because the singular limit:
\be
\lim_{\varepsilon\downarrow 0} e^{-iH_0t}\left(
-\Delta-(\lambda_0+\mu)\mp i\varepsilon\right)^{-1}
\nn\ee
satisfies a local energy decay estimate as $t\to\pm\infty$; see    
{\bf (H4)}. Thus, for $t<0$, in
(\ref{eq:model-damping}) we replace $\Gamma$ by $-\Gamma$.
\medskip

  \nit{\it Normal Forms for coupled oscillator - wave systems}

  The damped oscillator above is a normal form equation, which  illustrates 
   a more general phenomenon. 
  Dynamical systems like (\ref{eq:model}) can be
 viewed as a Hamiltonian systems consisting of two subsystems:
 a single (or, in general, multiple)  degree of
freedom oscillator and an  infinite dimensional radiating wave field. 

\nit (I)  In (\ref{eq:model}) the discrete oscillator frequency does not
lie in  the  continuous spectrum of wave frequencies. However, the
time periodicity of the potential generates frequencies in the solution
which resonate with continuum modes.

\nit (II) In \cite{kn:TDRT0}, \cite{kn:TDRT1}, quantum resonance
type problems are  considered in which the discrete oscillator frequency
coincides with a frequency of continuum modes. Coupling of the two
occurs due to a time-independent localized potential.

\nit (III) In  our study of radiation damping of 
 "breather-like" structures \cite{kn:rdamping} we find, 
 as in (\ref{eq:model}), that  the discrete
oscillator frequency does not lie in  the  continuous spectrum of wave
frequencies. Here, the coupling is due to a 
nonlinear potential which generates frequencies which lie in the
continuous spectrum.

  In an isolated 
   single degree of freedom Hamiltonian system, the general normal
  form is:
  \be
  A'\ =\ i\left(c_{10}\ +\ c_{21}|A|^2\ +\ c_{32}|A|^4\ +\ ...\ 
   +\ c_{n+1,n}\ |A|^{2n}\ +\ ...\right)\ A,
   \label{eq:single-dof}
   \ee
   where the constants $c_{n+1,n}$ are {\bf real} numbers.

\nit{\it How is this normal form altered as a result of coupling of the
 oscillator to an infinite dimensional radiative wave equation ?}

In the \cite{kn:TDRT0}, \cite{kn:TDRT1}, \cite{kn:rdamping} and in the
present work, we find the more general {\it dispersive or radiative
Hamiltonian normal form}:
 \be
   A'\ =\ \left(k_{10}\ +\ k_{21}|A|^2\ +\ k_{32}|A|^4\ +\ ...\
	  +\ k_{n+1,n}\ |A|^{2n}\ +\ ...\right)\ A,
		 \label{eq:complexsingle-dof}
			\ee
with complex coefficients; 
\be k_{n+1,n}\ =\ d_{n+1,n}\ +\ ic_{n+1,n},\nn\ee
where $d_{n+1,n}$ and $c_{n+1,n}$ are real numbers.
Non-zero {\it real parts} of the normal form coefficients arise due to
coupling of discrete and continuum modes. In the linear problems: (I) and
(II), we always have $d_{10}\equiv -\Gamma\le0$. 
In  (III) we have $k_{10}=ic_{10}$ and, 
 depending on the details of the nonlinear interaction, we
 may have: $k_{21}=ic_{21}$ and $k_{32}=-\Gamma +ic_{32}$
 \cite{kn:rdamping}, $k_{21}=-\Gamma+ic_{21}$ \cite{kn:PSW}, or other
 possibilities. 
 In each case there's a different expression for $\Gamma$, the first
  nontrivial {\it real} contribution to a normal form coefficient, but
   the
	essence of its calculation is as described above.
 Generically one has $-\Gamma<0$, and it is natural to conjecture that the
 first non-zero $\Gamma=d_{n_*+1,n_*}$ is strictly negative.

\bigskip

\nit{\it Connection with persistence theory of periodic solutions:}
\medskip

The condition $\Gamma > 0$ (see also {\bf
(H6)}) can be seen to arise in another manner.  Motivated by the
perturbative approach to constructing periodic solutions of ordinary
differential equations, it is natural to seek a solution of
(\ref{eq:model}) in the form:
\be
\phi(t)\ =\ e^{-i\lambda_0 t}\psi_0\ +\ \psi_1\ +\ \psi_2\ +\ ...\ +\
\psi_j\ +\ ...,\label{eq:ansatz1}
\ee 
where $\psi_j$ is formally of order $\beta^j$. Substitution  of the
ansatz (\ref{eq:ansatz1}) into (\ref{eq:model}) yields that:
\be \psi_1\ =\ \psi_1^a\ +\ \psi_1^b,\label{eq:psi1}\ee
where
\ba
i\D_t\psi_1^a\ &=&\ H_0\ \psi_1^a\ +\ {1\over2}\
  e^{-i(\lambda_0-\mu) t}
 \beta\psi_0,\nn\\
 i\D_t\psi_1^b\ &=&\ H_0\ \psi_1^b\ +\ {1\over2}\
e^{-i(\lambda_0+\mu) t}\ \beta\psi_0.\
 \ea
These equations can be solved by setting:
\be
\psi_1^a\ =\ e^{-i(\lambda_0-\mu)t}\ \Psi^a,\ \psi_1^b\ =\
e^{-i(\lambda_0+\mu)t}\ \Psi^b.
\nn\ee
For $\psi_1^a$ we obtain the equation:
\ba
\left(\ H_0\ -\ (\lambda_0 -\mu)\ \right)\ \Psi^a\ &=&\
{1\over2}\beta\psi_0\nn\\
 \left(\ H_0\ -\ (\lambda_0+\mu)\ \right)\ \Psi^b\ &=&\
{1\over2}\beta\psi_0
\label{eq:ab}
\ea
The first equation of (\ref{eq:ab}) has a localized solution because
$\lambda_0-\mu<0$ and lies in the resolvent set of $H_0$.
Note however, that $\lambda_0+\mu>0$, and therefore
 the second equation can have a localized solution only if 
the projection of $\beta\psi_0$ onto the generalized
eigenfunction associated with the continuum energy $\lambda_0+\mu$
vanishes. This is equivalent to 
 $\Gamma=0$, where $\Gamma$ is  defined by (\ref{eq:model-fgr}); See also 
  (\ref{eq:model-fgr-ft}) or, more generally, {\bf (H6)}.
 Generically, 
 $\Gamma>0$.  The results of this
paper show, in particular, that this obstruction to solvability in the
class of quasiperiodic functions, implies radiative decay of solutions of
the initial value problem.

\bigskip
Finally, we discuss various contexts, where problems
involving time-dependent potentials have been studied.
 Equations with time dependent potentials 
arise in mathematical physics and are important in applications, {\it
e.g.} electron microscopy and solid state devices.
In quantum physics their analysis dates back to the early days of the 
theory, where the process of ionization by external field (light waves) was 
studied and later nuclear spin resonance and charge transfer problems; 
 see e.g. \cite{kn:L-L}, \cite{kn:G-P}. The problem of ionization and 
 more generally of excitation of a molecule by radiation is also 
 important in laser optics and chemistry.  Recent models of 
  chaos \cite{kn:J-L} and stochastic resonances also involve time dependent 
  potential models, {\it e.g.} \cite{kn:V-R}. The expression
	(\ref{eq:model-fgr}) (see also {\bf (H6)})
   is the analogue of the Fermi Golden Rule, which
	 arises in the study of spontaneous emission,
   autoionization \cite{kn:G-P}, \cite{kn:L-L}, \cite{kn:RS4}. 
	In the context of the
	   problem of
		 ionization of an atom by a time-dependent electric field a
		 heuristic
		   derivation is presented
				 in section 42 of  \cite{kn:L-L}.

%
%

  Equations of type (\ref{eq:abstracttdse}) and (\ref{eq:tdse}) are also 
   relevant   
  in the study of coherent 
  solutions of nonlinear systems. For example, a periodic
  or quasiperiodic solution can be viewed as a bound state of an
  equation of this type  with a self-consistent (nonlinear)
  time-dependent potential. This point of view was first exploited by
  Sigal \cite{kn:sigaljapan}, \cite{kn:PS} who 
  used the methods of 
   dilation analyticity to relate the generic structural instability of
   quasi-periodic solutions of certain nonlinear systems to the generic
   instability of an embedded eigenvalue in the continuous spectrum of a
   Floquet operator. This approach
   yields the spectral line shift and the Fermi golden rule, which
   give the location of the {\it resonance} to which the embedded eigenvalue
   has been perturbed. 
  The approach we have taken to the non-persistence question
   is to show by a study of the large time behavior
  of the initial value problem that solutions decay slowly as
  $t\to\infty$. 

The analysis of time dependent resonant 
 problems in the physics literature is based on 
formal time dependent perturbation theory; see 
 \cite{kn:L-L}, \cite{kn:G-P}, where Dyson series to first order is
 used, and \cite{kn:Sa} where Dirac's perturbation theory is applied.
This analysis is limited in many respects. 
 In \cite{kn:L-L}, \cite{kn:G-P} it is  assumed, among other things, that the 
relevant time scale of the approximation is smaller than $\ve^{-1}$
where $\ve$ is the perturbation size, while we know that the 
lifetime, is of order $\ve^{-2}$;  It is also not clear how to get the 
spectral line shift from this approach. 

Problems involving time dependent potentials have received considerable
attention in the mathematical physics literature during the last twenty
years, and a number of different approaches, based on scattering theory
techniques, have been developed.
In \cite{kn:How 1}  a kind of Floquet theory is used to analyze the case of 
 time-dependent potentials  decaying like $L^1$ functions in time. 
 The approach has also been used to treat some problems
  involving time-periodic
 potentials and long range potentials \cite{kn:Kit-Ya},
 \cite{kn:Ya}, \cite{kn:How 2}. 

 Another type of time-dependent problem is the charge-transfer problems.
 These are problems in which the potential $W$ is a sum of terms of the
 type: $W(x-vt)$. In \cite{kn:Yaj-chargetransfer}   
  a charge transfer problem is mapped in a 
 similar way onto a three-body type problem    
which is treated by resolvent methods; 
In \cite{kn:E-V} a geometric approach to scattering was used 
 to study the general behavior of scattering states with time-dependent potentials. 

For potentials decaying slowly in time which are perturbations of 
 many body hamiltonians, 
which appear in the analysis of N-body long range scattering, 
phase-space methods were used in  \cite{kn:Sig-Sof-Nbody}, 
 based on the 
 existence of an asymptotic energy operator in such cases.
 The AC Stark effect was treated in \cite{kn:Yaj-stark} by methods
exploiting the special form of the potential.
 Also related to our study is research concerning the wave equation with a time
periodic potential or the free wave equation in the exterior of a
periodically oscillating obstacle; see, for example, 
 \cite{kn:CS}, \cite{kn:Vain}, and those cited therein.

 We consider a Hamiltonian, $H_0$, with one bound state, 
 perturbed by a small and localized potential, 
 time-periodic potential of the simple form given in Hypothesis {\bf (H5)}. 
This assumption is made to simplify computations but the general method
applies to problems with more than one bound state and 
 to more general time - periodic or quasi-periodic  
 perturbations.
An application currently under investigation involves the
structural instability and  meta-stability of "breather-like" modes of 
certain linear wave-guide problems related to "M-soliton" solutions of
integral nonlinear flows \cite{kn:MSW}. 

In this paper we treat a model of the ionization problem \cite{kn:L-L},
\cite{kn:G-P}.  
By treating the process of ionization as a resonance caused by 
coupling of the point spectrum to the continuous spectrum, we apply the 
time dependent resonance theory, recently developed in 
\cite{kn:TDRT0}, \cite{kn:TDRT1} to find the large time behavior 
 of the solution.  
In particular, we show that 
under our conditions, the bound state is always disintegrating and 
 find the lifetime, $\tau\sim\Gamma^{-1}$,  and the 
 spectral line shift, (\ref{eq:rhoexpansion}); see the remark following
 Theorem 2.2.
  We also prove local decay estimates for the solution, which are of 
general interest to nonlinear and other problems.
\bigskip
\vfil\eject
\nit {\bf Notations and terminology:} 

\nit Throughout this paper we will denote a generic
constant by $C$, $D$, etc. 

\nit $\langle x\rangle\ =\ \left(
1+|x|^2 \right)^{1\over2}$.

\nit ${\cal L}({\cal H}) =\ $ the space of bounded linear operators on ${\cal H}$

\nit Functions of self-adjoint operators are defined via the spectral theorem
\cite{kn:RS1}.

\bigskip

\nit{\bf Acknowledgements:} 
The authors wish to thank
 Eduard Kirr for helpful comments on the manuscript. AS and MIW are
supported in part by grants from the National Science Foundation.

\section{General formulation and main results.}

Consider the general system
\be
i \D_t\phi(t)\ =\ \left(H_0 + W(t)\right)\phi(t) , \label{eq:generalse}
\ee
Here, $\phi(t)$ denotes a function of time, $t$, with 
values in a complex Hilbert space ${\cal H}$. 

\nit\underline{\bf Hypotheses on $H_0$:}

\nit{\bf (H1)} $H_0$ is self-adjoint 
on ${\cal H}$ and both $H_0$ and $W(t),\ t\in\R^1,$ are 
 densely defined on a subspace ${\cal D}$ of ${\cal H}$. 
 
\nit The norm on
 ${\cal H}$ is denoted by $\|\cdot\|$, and the inner product of $f,g\in
 {\cal H}$, by $\left(f,g\right)$.

\nit{\bf (H2)} 
The spectrum of  $H_0$  is assumed to consist of an absolutely continuous part, $\sigma_{\rm cont}(H_0)$, with associated spectral projection 
 $\Pc$ and a  single isolated eigenvalue $\lambda_0$ 
 with corresponding normalized eigenstate, $\psi_0$, {\it i.e.}
\be H_0\psi_0\ =\ \lambda_0 \psi_0,\ \| \psi_0\|=1.\ee

The manner in which we shall measure the decay of solutions is 
 typically in a local decay sense, {\it e.g.} 
 for the scalar Schr\"odinger equation we measure local decay using
 the norms:
$f\mapsto \| \langle x\rangle^{-\sigma} f \|_{L^2}$, where $\sigma>0$. 
 So that our theory applies  to a class of general systems (involving,
 for example, 
 vector equations with matrix operators), we assume the existence of 
 self-adjoint "weights", $\wmi$ and $\wpl$ such that 

\ (i)\ $w_+$ is defined on a dense subspace of ${\cal H}$ and on 
 which $w_+\ge cI$,
 \ $c > 0$.

\ (ii)\  $w_- \in {\cal L}({\cal H})$

\ (iii)\ $\wmi\ \wpl \Pc\ =\ \Pc\ =\ \Pc\ \wmi\ \wpl$

In the scalar case $\wpl$ and $\wmi$ 
 correspond to multiplication by 
  $ \langle x\rangle^\sigma$ and    $\langle x\rangle^{-\sigma}$, 
 respectively; see section 3.

The following hypothesis and Corollary ensure that the equation
satisfies sufficiently strong dispersive time-decay estimates.

\nit {\bf (H3)} \underline{Local decay estimates on $e^{-iH_0t}$}:
Let $r\ge2+\varepsilon$, where $\varepsilon>0$.
 There exist $w_+$ and $w_-$, as above,  such that 
for all $f$:

\nit 
\be
{\rm {\bf (a)}}\ \ \|  \wmi e^{-iH_0t} {\bf P_c} f \| \le C\ \lan t\ran^{-r+1} \|\wpl  f \|,
\label{eq:localdecay}
\ee
\nit and
 \be
{\rm {\bf (b)}}\ \ \|\wmi e^{-iH_0t}(H_0-\lambda_0 - \mu -i0)^{-1}\
 \Pc f\| \le
 \ C\lan t\ran^{-r+1}\ \|\wpl f\|.
  \label{eq:singularldest}
\ee

\nit (1) Typically one is not "handed" the singular estimate (\ref{eq:singularldest}).
It can however be proved using (\ref{eq:localdecay}) and the following: 
\medskip

\nit {\bf (b$_{\Delta}$)} 
 Let  $\Delta$ denote an open interval containing $\lambda_0+\mu$  
 and contained in the interior of the 
  continuous spectrum of $H_0$. Let ${\bf g_\Delta}$
 denote a smooth function of compact support which is identically one on 
 $\Delta$. Then, for all $f$ such that $\wpl f\in {\cal H}$
\be
\|\wmi e^{-iH_0t} {\bf g_\Delta}\ f\| \le C_\Delta\ \lan t\ran^{-r} 
\|\wpl f\|.\label{eq:localdecay1}
\ee
\medskip

\nit (2) To prove that  (\ref{eq:singularldest}) is implied by estimates
(\ref{eq:localdecay}) and (\ref{eq:localdecay1}), one can follow 
very closely the approach taken in \cite{kn:TDRT0}, \cite{kn:TDRT1}.  
In particular,  
see Proposition 2.1 and  Appendix A of \cite{kn:TDRT1}. 
In \cite{kn:TDRT1},  the proofs are carried out in for a particular
choice of ${\cal H},\ \wpl,$ and  $\wmi$, but as indicated in the
remark following the main theorem of \cite{kn:TDRT1}, the proof can be
adapted to a more general setting and is  virtually unchanged; one
need only replace $L^2$ by a general Hilbert space ${\cal H}$, $\lan
x\ran^{-\sigma}$ by $\wmi$, and $\lan x\ran^\sigma$ by $\wpl$.
 This requires hypothesis:

\nit{\bf (H4)}  There is a choice of real number $c$ such that
\be
\|\wpl (H_0+c)^{-1}\wmi\|_{{\cal L}({\cal H})}
\label{eq:H4}
\ee
can be made sufficiently small.

\nit
In the scalar case $\wpl$ and $\wmi$
 correspond to multiplication by
   $ \langle x\rangle^\sigma$ and    $\langle x\rangle^{-\sigma}$,
	respectively; see section 3.

\nit {\bf (H5)} \underline{\bf Hypotheses on the perturbation $W(t)$:}

We consider  
 time-dependent symmetric perturbations of the form
\be
W(t) =  \cos \mu t\  \beta,\ {\rm with}\ \beta^*=\beta.
\label{eq:generalW}
\ee
In many applications, $\beta$, is a spatially
localized scalar or matrix function. 

To measure the size of the perturbation  $W$, we introduce the norm
\be
||| W |||\ \equiv\ \left(  \| \wpl\; \beta\|^2_{{\cal L}({\cal H})}
\ +\  ( \| \wpl\; \beta\wpl\|^2_{{\cal L}({\cal H})}\right)^{1\over2}.
\label{eq:betanorm}
\ee

Our final hypothesis is a resonance condition which says that 
 $\mu
+\la_0\ \in \sigma_{\rm cont}(H_0)$ and that there is nontrivial
coupling.

\nit {\bf (H6)} \underline{Resonance condition}
There exists $\theta_0>0$, independent of $W$,  such that 
\be
\Gamma\ \equiv\ {\pi\over4}\ 
  \left( \beta\psi_0, \; \delta(H_0 - \la_0 - \mu) \beta\psi_0
\right) \ge
 \theta_0||| W |||^2 > 0
 \label{eq:tdfgr}
 \ee

\nit{\bf Remark:} 
 That the expression in  (\ref{eq:tdfgr}) and the principal
value term in (\ref{eq:pvterm}) below are finite is a consequence \cite{kn:BFSS}
 of the
local decay estimate {\bf (H3)} and the representation:
\be
\left(\ \beta\psi_0,\ (H_0-\lambda-\mu-i0)^{-1}\ \beta\psi_0\ \right)\ 
 =\ i\ \int_0^\infty\ \left(\ \beta\psi_0,e^{-i(H_0-\lambda_0-\mu)s}
 \beta\psi_0\ \right)\ ds.
 \nn\ee

Our main result is the following:
\begin{theo}
 Let $H_0$ satisfy hypotheses {\bf (H1)-(H6)} and assume
 $w_+\phi_0\in{\cal H}$. 
Then,  if $||| W |||$ is sufficiently small, 
 the solution of (\ref{eq:generalse}) satisfies the local decay
estimate:
\begin{equation}
\| \wmi\ \phi(t)\| \le C\lan t\ran^{-r+1} \| \wpl\ \phi_0\|,\ \ t\in\R.
\label{eq:ldestimate}
\end{equation}
\end{theo}

\nit{\bf Remarks:} 
\ In  the proof we actually use a less stringent property than {\bf (H6)}.
 Rather, we use that:
  
 \nit{\bf (H6')} $|||W|||^3\ \Gamma^{-1}$ is sufficiently small.

Under the same hypotheses as Theorem 2.1, we obtain   
 more detailed information on the behavior of $\phi(t)$: 

\begin{theo} 
\begin{eqnarray}
\phi(x,t) &=& a(t)\psi_0 + \phi_d(t),\ \ \left(\psi_0\ ,\ \phi_d(t)\right)\ =\ 0,\nn\\
a(t) &=& a(0)\ e^{-\Gamma t}\ e^{-i\omega_1(t)}\  +\ R_a(t)\nn \\
\phi_d(t) &=& e^{-iH_0t}\ \Pc \phi_0\ +\ {\tilde \phi(t)}.\nn
\end{eqnarray}
For any fixed $T_0>0$, we have 
\be
\left| R_a(t)\right|\ \le\ C\ |||W|||,\ |t|\le T_0 \Gamma^{-1}
 \label{eq:Raestimate}\ee
 Moreover,
 \ba
 |R_a(t)| &=& {\cal O}(\lan t\ran^{-r+1}),\ |t|\to\infty\nn\\
\omega_1(t)\ &=&\ \lambda_0\ t\  
-\ {1\over4}\left(\beta\psi_0,(H_0-\lambda_0+\mu)^{-1}\Pc\beta\psi_0\right)\
 t\ \nonumber\\
 &&\  -\ {1\over4}\left(\beta\psi_0,\ {\rm P.V.} 
  (H_0-\lambda_0-\mu)^{-1}\ \Pc\beta\psi_0\right)\
   t\ +\ M(t) ,
  \label{eq:pvterm}\ea
  where $M(t)$ is of order $|||W|||^2$ and is oscillatory in $t$.
 Furthermore, ${\tilde \phi}= \phi_1 + \phi_2$ is  explicitly given in 
 (\ref{eq:intphideqn}), with 
 $\| w_-\tilde\phi(t)\|\ =\ {\cal O}(\lan t\ran^{-r+1})$ as $|t|\to\infty$, and therefore,
 by {\bf (H3)} same holds  $\| w_-\phi_d(t)\|\ =\ {\cal O}(\lan t\ran^{-r+1})$ as
 $|t|\to\infty$.
\end{theo}
\medskip

\nit{\bf Remark:}  
Suppose the initial data is given by the bound state of the
unperturbed problem, {\it i.e.} $\phi(x,0)=\psi_0(x)$, $a(0)=1$,\ $\phi_d(0)=0$.
 Then, from the
expansion of the solution 
 we have that for  $0\le t\le \Gamma^{-1}$ that 
 the projection of the solution on 
$\psi_0$ is of order $e^{-\Gamma t}$, with an error of order $|||W|||$.
Hence it is natural to view the state $\psi_0e^{-i\lambda_0 t}$ as a
{\it metastable state}, with lifetime $\tau =\Gamma^{-1}\sim
 |||W|||^{-2}$.

\section{ An example: the Schr\"odinger equation }

In this section we apply the results of Theorems 2.1 and 2.2 to the
special case of the Schr\"odinger equation with a   
time-periodic  and spatially localized perturbing potential, equation
(\ref{eq:model}). 
Models of the sort considered in this example occur in the study of
ionization
of an atom by a time-varying electric field; see \cite{kn:L-L},
\cite{kn:G-P}.

We begin by indicating assumptions on $V(x)$ and $\beta(x)$ under which
{\bf (H1)-(H6)} hold.

We take ${\cal H}=L^2(\R^n)$, and 
 $H_0\ \equiv\ -\Delta\ +\ V(x)$, where $V(x)$ and $\beta(x)$ are real-valued, 
sufficiently differentiable and decaying rapidly as $x\in\R^n$ tends
 to infinity. Thus, $H_0$ is self-adjoint,  
  and $H_0$ and $W(t)$ are densely defined operators in
 $L^2$.
Moreover, we assume that $H_0$ has exactly one negative eigenvalue 
 and the absolutely 
 continuous spectrum of $H_0$
is equal 
to the positive half-line.    

With regard hypothesis  {\bf (H3)}, we take the weights used to measure 
 local energy decay  
 to be  $w_\pm \equiv \lan x\ran^{\pm\sigma}$.
There is a good deal of literature on local energy decay estimates for 
$e^{-iH_0t}\Pc$. These results
require sufficient regularity and decay of the potential $V(x)$.
 We refer the reader to \cite{kn:JK}, \cite{kn:JSS} and \cite{kn:Murata};
 see also \cite{kn:Rauch}, \cite{kn:Schonbek}.

The large time expansion of the operator: $\lan
x\ran^{-\sigma}e^{-iH_0t}\Pc\lan x\ran^{-\sigma}$ in $L^2$ contains a
{\it non-generic} leading order term of order $t^{-{1\over2}}$.  More
rapid decay as $t\to\infty$ is ensured provided we assume the following
 condition concerning the behavior of the resolvent of $H_0$ near  zero energy:

{\bf (Z)}  Zero is not
a resonance of the operator $-\Delta+V(x)$. For a precise definition of
{\it zero energy resonance} in this context and a  detailed discussion
 see \cite{kn:JK} and 
\cite{kn:Murata}. 

\nit{\bf Remark:} The condition {\bf (Z)} holds generically in the space
of potentials.

Hypothesis {\bf (H3a)} is a consequence of:

\begin{prop} Assume that $V(x)$ is sufficiently regular and decays
sufficiently rapidly as $|x|\to\infty$; see \cite{kn:JK}, \cite{kn:JSS},
\cite{kn:Murata}. Additionally, assume condition {\bf (Z)}. Then,  
we have the following local energy decay estimate.

\nit (a) Assume  $n\ge3$ and $\sigma>1$.
 Then, there exists $r>2$ such 
that
\be \| \lan x\ran^{-\sigma}\ e^{-iH_0t}\Pc f\ \|_2\ \le\ 
C\ \lan t\ran^{-r+1}\ \|\lan x\ran^\sigma f\|_2.
\label{eq:ldest}\ee

\nit (b) Assume $n=1$. Then, we get (\ref{eq:ldest}) with 
 $\lan t\ran^{-r+1}=\lan t\ran^{-{3\over2}}$.

\nit (c) Assume $n=2$. Then, we get (\ref{eq:ldest}) with 
 $\lan t\ran^{-r+1}$ replaced by 
 $\lan t\log^2t\ran^{-1}$.
\end{prop}

\medskip

Hypothesis {\bf (H3b$_\Delta$)} is a consequence of:

\begin{prop} Assume the above hypotheses on $H_0$ and assume that 
$\Delta$ denotes an  open and bounded subinterval of the continuous spectrum
of $H_0$. Let $g_\Delta$ denote a smooth characteristic function, as defined in 
section 2. Then, for $2<r<\sigma$, 
\be
\|\lan x\ran^{-\sigma}\ e^{-iH_0t}g_\Delta(H_0) f\ \|_2\ \le\ 
C\ \lan t\ran^{-r}\ \|\lan x\ran^\sigma f\|_2.
\no\ee
\end{prop}
\nit{\it proof:} If $H_0$ is replaced by the Laplacian, this estimate is 
straightforward since $ g_\Delta(H_0) f$ is spectrally supported away from
points of stationary phase. An approach to this estimate for $H_0$, a variable
coefficient operator is reviewed in Appendix A and is  based on the 
{\it Mourre estimate}, where the operator $A$ can be taken to be: $A=(x\cdot
p+p\cdot x)/2$, where $p=-i\nabla$.

We take the time-dependent perturbation to be of the form:
\be 
W(x,t) \ =\ \cos{\mu t}\ \beta(x),
\no\ee
with $\beta$ sufficiently differentiable and rapidly decaying in $x$, {\it e.g.}
 $\lan x\ran^{2\sigma}\beta(x)\in L^2$. Thus, {\bf (H5)} is satisfied as well. 
  Finally, the resonance condition  
{\bf (H6)}, holds generically in the potential $V(x)$.

Therefore, our main theorems  
Theorems 2.1  and Theorems 2.2 on the  structural instability of the
unperturbed bound state, and large time behavior of such systems apply.
\section{ Decomposition and derivation of the dispersive normal form}

 As in \cite{kn:TDRT0}, \cite{kn:TDRT1}, we begin by deriving a decomposition of 
 the solution, $\phi(t)$
 , which will facilitate the study of its large time behavior.
 Let  
 \begin{equation}
\phi(t) = a(t) \psi_0 + \phi_d(t), \label{eq:ansatz}
\end{equation}
with the orthogonality condition
\begin{equation}
(\psi_0, \phi_d) = 0 \ \ {\rm for \ all} \ \ t \label{eq:orthog}.
\end{equation}
Note therefore that $\phi_d\ =\ \Pc\phi_d$ since $H_0$ has been assumed
to have only one eigenvalue.

\nit{\bf Remark:} The required modifications in our analysis for the case 
  where $W(t)$ has more general periodic or 
quasi-periodic time dependence, and when $H_0$ has more than one bound state can
be seen at this point. If $H_0$ has $M$ discrete eigenvalues, then the decomposition
 (\ref{eq:ansatz}) is replaced by a sum over the $M$  discrete modes with
 amplitudes $a_j(t),\ j=0,...,M-1$, plus the continuous 
 spectral part. If $W(t)$ has more general periodic or quasi-periodic time dependence,
 in the analysis of the coupled equations for $a_j(t)$ and $\phi_d(t)$ (see below), we 
 use an expansion of $W(t)$ in terms of operators of the form (\ref{eq:generalW}). 
\medskip

We proceed by first inserting (\ref{eq:ansatz}) into (\ref{eq:generalse}),
which 
yields the equation:
\ba
i\D_ta(t)\psi_0\ +\ i\D_t\phi_d(t)\ &=&\ \lambda_0a(t)\psi_0\ + H_0\phi_d(t)
\no\\
&+&\ a(t)W(t)\psi_0\ +\ W(t)\phi_d(t)\label{eq:substit}
\ea

Taking the inner product of (\ref{eq:substit}) with  $\psi_0$ yields the 
  following equation for $a(t)$:
\ba
i \D_t a\ &=&\ \lambda_0 a(t)\ +\  \left(\psi_0, W(t)\psi_0\right) a(t)\ +\
 \left(\psi_0, W(t)\phi_d\right),\label{eq:aeqn1}\\
 a(0)\ &=&\ \left(\psi_0,\phi(0)\right)\no
\ea
In deriving (\ref{eq:aeqn1}) we have used that $\psi_0$ is normalized and
the relation
\be \left(\psi_0 , \D_t\phi_d\right) = 0,\nonumber\ee
a consequence of (\ref{eq:orthog}).

Applying $\Pc$ to (\ref{eq:substit}), we obtain an equation for
$\phi_d$:
\ba
i \D_t \phi_d(t)\ &=&\ H_0 \phi_d(t)\ +\ \Pc   W(t)\phi_d(t)\ +\ a(t)\Pc
W(t)\psi_0,\label{eq:phideqn}\\
\phi_d(0)\ &=&\ \Pc\phi(0)\no
\ea
Since we are after a slow resonant decay phenomenon, it will prove
advantageous to extract the fast oscillatory behavior of $a(t)$. 
We therefore define:
\be    A(t) \equiv e^{i\la_0 t} a(t).\label{eq:Adef}\ee
Then, (\ref{eq:aeqn1}) reads
\begin{equation}
\D_tA\ =\  -iA\left(\psi_0, W(t)\psi_0\right)\ -\ ie^{i\la_0 t}\ 
 \left(\psi_0, W(t)\phi_d(t)\right).\label{eq:Aeqn1}
\end{equation}
Solving (\ref{eq:phideqn}) by Duhamel's formula we have
\ba
\phi_d(t) &=& e^{-iH_0t}\phi_d(0)\ -\ 
 i \int^t_0 e^{-iH_0(t-s)}  \Pc W(s) a(s) \psi_0 ds \nonumber \\
&-&   i \int^t_0 e^{-iH_0(t-s)} \Pc  W(s) \phi_d(s)\ ds\no \\
&\equiv& \phi_0(t)\  +\ \phi_1(t)\  +\ \phi_2(t) \label{eq:intphideqn} .
\ea
By standard methods, the system 
 (\ref{eq:Aeqn1})-(\ref{eq:intphideqn}) for $A(t)$
and $\phi_d(t) = \phi(t) - e^{-i\lambda_0t}\ A(t)\ \psi_0$
has a global solution in $t$ with 
$$A\in C^1(\R),\  
 \|\phi_d(t)\|\in C^0(\R),\  \|w_-\phi_d(t)\|\in C^0(\R).
$$
 Our analysis of the $|t|\to\infty$ behaviour is based on a study of
 this system.

We next insert (\ref{eq:intphideqn}) into (\ref{eq:Aeqn1}) we get
\begin{equation}
\D_tA(t)\ =\ -iA(t) \left(\psi_0, W(t) \psi_0\right)\ -\ 
 ie^{i\la_0t} \sum^2_{j=0} \left(\psi_0, W(t)\phi_j\right).\label{eq:Aeqn2}
\end{equation}

We next give a detailed expansion of the sum in
(\ref{eq:Aeqn2}). It is in the  $j=1$ term
  that the key
resonance is found. This makes it  possible to find a normal form for
(\ref{eq:Aeqn2}) in which {\it internal damping} in the system 
 is made explicit. This damping is responsible for the energy transfer
 from the discrete to continuum modes of the system and the associated
 radiative decay of solutions.

\begin{prop}
\begin{eqnarray}
& &-ie^{i\la_0t} \cos \mu t \left(\ \beta\psi_0, \ \phi_1(t)\  \right)\ =
\  R_1 + R_2 + R_3 + R_4,\ {\rm where}\  \nonumber \\
R_1 &=& -{1\over2} e^{i\lambda_0t}\cos{\mu t}
 \int^t_0 \left( \beta\psi_0, e^{-i H_0(t-s)} 
 \  e^{i(\mu - \la_0)s} A(s) \Pc  \beta \psi_0 \right) ds \no\\
R_2 &=& {i \over 2} e^{-i\mu t} \cos{\mu t}\ A(t)\ 
 \left( \beta\psi_0, (H_0-\la_0-\mu -i0)^{-1} \Pc  \beta\psi_0 \right) \no
 \\
R_3 &=& - {i\over 2} A(0) e^{i\lambda_0 t} \cos{\mu t} 
 \left( \beta\psi_0, e^{-iH_0t} (H_0-\la_0-\mu -i0)^{-1} \Pc  \beta \psi_0 
  \right) \nonumber \\
R_4 &=& -{i\over 2}  e^{i\la_0 t} \cos \mu t \int^t_0 
 \left( \beta\psi_0, e^{-iH_0(t-s)} (H_0-\la_0-\mu -i0)^{-1} 
  e^{-i(\la_0 + \mu)s} \D_sA(s) \Pc  \beta\psi_0 \right) ds.
  \nonumber\\
  \label{eq:R1234}
\end{eqnarray}
\end{prop}

\noindent
\underline{Proof of Proposition 4.1}  

Using that $\cos{\mu t} = {1 \over 2} (e^{i\mu t} + e^{-i\mu t})$ and
the definition $A(t)=e^{i\lambda_0t}a(t)$, we get from (\ref{eq:intphideqn})
\ba
\phi_1(t) = &-{i\over 2}& \int^t_0 e^{-iH_0(t-s)} 
 e^{-i(\la_0 - \mu)s }A(s) \Pc  \beta\psi_0\ ds \no \\
&-{i\over 2}& \int^t_0 e^{-iH_0(t-s)} e^{-i(\la_0 + \mu)s} A(s) 
 \Pc  \beta\psi_0 ds\no\\
&\equiv&\  K_0+K_{\rm res}.\label{eq:phidexpand}
\ea
Therefore,
\ba
-ie^{i\la_0 t}\ \cos{\mu t}\left(\beta\psi_0,\phi_1(t)\right)\ &=&\ 
 -ie^{i\la_0 t} \cos{\mu t}\left(\beta\psi_0,K_0+K_{\rm res}\right)
 \nonumber\\
 &=&\ R_1\ -\ ie^{i\la_0 t}\cos{\mu t}\left(\beta\psi_0, K_{\rm res}\right)
\label{eq:jeq1expand}
\ea
The terms $R_j,\ j=2,3,4$ in (\ref{eq:R1234}) come from an
expansion of the last term in (\ref{eq:jeq1expand}). This is the source
of the 
 key resonant contribution. 
 We now proceed with a careful
evaluation of this term. 

We first regularize $K_{\rm res}$.
For $t>0$, let $\ve$ be positive and arbitrary and define:
\be
K^\ve_{\rm res}\ =\ -{i\over 2} \int^t_0 e^{-iH_0(t-s)} 
 e^{-i(\la_0 + \mu -i\ve )s} A(s)\ 
 \Pc\  \beta\psi_0\ ds
\label{eq:Kepsilon}
\ee
Note that $K_{\rm res} = \lim_{\ve\to0}K^\ve_{\rm res}$. 
Integration of the expression in (\ref{eq:Kepsilon})  
by parts, and letting $\ve$ tend to zero from above gives

\begin{prop}
The following expansion of $K_{\rm res}$ is valid in ${\cal S}'$:
\ba
K_{{\rm res}} &=& - {1 \over 2} \ e^{-i(\lambda_0 +\mu)t} A(t) 
  (H_0 - \la - \mu- i0)^{-1} \Pc  \beta\psi_0 \no \\
&+& {1\over 2} A(0)\ e^{-iH_0t} (H_0 - \la - \mu -i0)^{-1} \Pc  \beta\psi_0
\no\\
&+& {1\over 2}\int_0^t e^{-iH_0(t-s)} 
 (H_0 - \la - \mu- i0)^{-1} e^{-i(\la_0+\mu)s} \D_sA(s) \Pc  \beta\psi_0 ds 
 \label{eq:Kresexpansion}
\ea
\end{prop}
The choice of regularization, $-i\ve$, in (\ref{eq:Kepsilon}) ensures
that the latter two terms in the expansion of $K_{\rm res}$,
(\ref{eq:Kresexpansion}), decay dispersively as $t \to +\infty$; see
hypothesis {\bf (H3)} and section 6. For $t<0$, we replace
$+i\ve$ with $-i\ve$ in  (\ref{eq:Kepsilon}).

The proof of Proposition 4.1 is now completed by substitution 
 of the expansion (\ref{eq:Kresexpansion}) for 
 $K_{{\rm res}}$ into the last term in (\ref{eq:jeq1expand}).
\ \ \ []
\bigskip

%
We further expand the terms $R_1$ and $R_2$ to make explicit the
resonant contributions.
\begin{prop} 
\ba
R_1\ &=&\ \rho_1(t)A(t)\ +\ R_{1a}(t)\ +\ R_{1b}(t),\label{eq:R1expand}\\
R_2\ &=&\ (\rho_2(t)\ -\ \Gamma ) A(t),\label{eq:R2expand}
\ea
where
\begin{eqnarray}
\rho_1(t)\ &=& {i\over4}(1+e^{2i\mu t})
 \left(\beta\psi_0,(H_0-\lambda_0+\mu)^{-1}\Pc\beta\psi_0\right)\no\\
 \rho_2(t)\ &=&\  {i\over 4}(1 + e^{-2i\mu t})
  \left(\beta\psi_0, {\rm P.V.}\ (H_0 - \la_0 - \mu)^{-1}\
   \Pc\beta\psi_0\right)\no\\
  \quad\quad &-&{\pi \over 4} e^{-2i\mu t}\
  (\beta\psi_0, \delta(H_0 - \la_0 - \mu) \Pc\beta\psi_0)
  \nonumber\\ 
\Gamma\ &=&\ {\pi \over 4}
 \left(\beta\psi_0, \delta(H_0 - \la_0 - \mu) \Pc\beta\psi_0\right)
   \nonumber \\
R_{1a}(t)\ &=&\ -{i\over2} A(0)e^{i\lambda_0 t}\cos{\mu t}\ 
 \left(\beta\psi_0,\ e^{-iH_0t}\ (H_0 - \la_0 + \mu)^{-1} 
 \Pc \beta\psi_0\right)\no  \\ 
R_{1b}(t)\ &=&\ {1\over 2i} e^{i\la_0t}\ \cos{\mu t}\  
 \left(\beta\psi_0, \int_0^t\ e^{-iH_0(t-s)}
  e^{-i(\la_0-\mu)s}\D_sA(s)(H_0 - \la + \mu)^{-1}\Pc\beta\psi_0\right)\no
\label{eq:R1ab}
\end{eqnarray}
\end{prop}

\noindent
\underline{Proof.}
To obtain (\ref{eq:R2expand}) we use the well known distributional identity
\be  (x - i0)^{-1} = P.V.\ x^{-1} + i \pi\delta(x). \no\ee  
Equation (\ref{eq:R1expand}) follows by integration by parts. \ \ \ \ []

Combining equation (\ref{eq:Aeqn2}) with the previous propositions,
 we have 

\begin{prop}
\be
\D_t A(t)\ =\ \left(\ -\Gamma\ +\  \rho(t)\ \right)\ A(t)\ +\ E(t), 
\label{eq:Aeqn3}
\ee
where 
\be
E(t)\ =\ 
  R_{1a} + 
 R_{1b} + R_3 + R_4 
  -ie^{i\la_0t}\cos \mu t \left(\beta\psi_0, \phi_0 (t)\right) \nonumber \\
- ie^{i\la_0 t} \cos \mu t\ \left(\beta \psi_0, \phi_2\right). 
\label{eq:Eexpand}
\ee
 Here, 
 $\phi_0$ and $\phi_2$ are given in (\ref{eq:intphideqn}), $R_{1a}$ and  
  $R_{1b}$ in (\ref{eq:R1ab}), $R_3$ and  $R_4$ in (\ref{eq:R1234}), and  
\ba \rho(t)\ &\equiv&\ 
 -i\cos{\mu t}\left(\psi_0,\beta\psi_0\right)\ +\  \rho_1(t) + \rho_2(t)
\no\\
&=&\
{i\over4}\left(\beta\psi_0,(H_0-\lambda_0+\mu)^{-1}\Pc\beta\psi_0\right)\no\\
&+&\
 {i\over4}\left(\beta\psi_0,{\rm P.V.}\ 
  (H_0-\lambda_0-\mu)^{-1}\Pc\beta\psi_0\right)\ +\ m(t),
  \label{eq:rhoexpansion}
\ea
where $m(t)$ is a bounded and oscillatory  function of zero mean. 
 \end{prop}
\noindent
{\bf Remarks:} 

\nit (1) The point of (\ref{eq:Aeqn3}) is that the damping coefficient
($\Gamma>0$), which arises due to the coupling of the discrete bound
state to the continuum modes by the periodic perturbation, is made explicit.
This facilitates a direct study of the transfer of energy from the bound
state to the continuum and the associated radiative
decay of solutions.

\nit (2) The leading order part 
 of equation (\ref{eq:Aeqn3}) is the analogue of the dispersive
normal form derived in \cite{kn:rdamping} for a class of nonlinear
dispersive wave
equations.

\bigskip

In the next section we  estimate the remainder terms in (\ref{eq:intphideqn})
and 
(\ref{eq:Aeqn3}).  

\section{ODE Estimates}

Equations (\ref{eq:intphideqn}) and (\ref{eq:Aeqn3}) comprise a dynamical
 system governing  
  $\phi_d(t)$ and $a(t)$, the solution of which is equivalent to the 
 original equation (1.1).  Our aim in this decomposition is to 
separate the dispersive (polynomially time-decaying) part $\phi_d(t)$ 
 from the solution, as well as 
 the transient (exponentially time-decaying) part.  
  In this and in the following section we derive a coupled system of
  estimates for $A(t)$ and $\phi_d(t)$, we show that 
  $A(t)$ decays in time, provided $\phi_d(t)$ is dispersively decaying
  and vice-versa.  Then, we exploit the assumption that the potential,
  $W$ is small to close the resulting inequalities, and prove the decay
  of both $A(t)$ and $\phi_d(t)$.  

  At this point, we state a simple Lemma which we shall use in a number of
  places in this and in the next section.

  \begin{lem}  Let $\alpha>1$.
  \be
  \int_0^t\ \lan t-s\ran^{-\alpha}\ \lan s\ran^{-\beta}\ ds\ \le
  C_{\alpha,\beta}\ \lan t\ran^{-\min (\beta, \beta +\alpha -1)}
  \nn\ee
  \end{lem}

  \nit\underline{Proof:} The bound is obtained by viewing the integral
  as decomposed into 
   a part over $[0,t/2]$ and the part over $[t/2,t]$.  We estimate
  the integral over $[0,t/2]$  by 
    bounding $\lan t-s\ran^{-\alpha}$ by its value at $t/2$ and explicitly
	computing the remaining integral. The integral over $[t/2,t]$ is
	computed by bounding $\lan s\ran^{-\beta}$ by its value at $t/2$ and
	again computing explicitly the remaining integral. Putting the two
	estimates together yields the lemma.

We now turn to the estimate for $A(t)$ in terms of the dispersive norm of 
 $\phi_d(t)$ and local decay estimates for $e^{-iH_0 t} \Pc  (H_0)$.

\begin{prop}
$A(t)$, the solution of (\ref{eq:Aeqn3}), can be expanded as:
\ba
  A(t)\ &=&\ e^{\int^t_0\rho(s)ds} \left( e^{-\Gamma t} A(0) + R_A(t) \right)
  \label{eq:prop4a}\\
  R_A(t)\ &=&\ \int_0^t\ e^{-\Gamma(t-\tau)}\ \tilde E(\tau)\ d\tau,
  \label{eq:prop4aa}
\ea
where $\tilde E(t)$ is given in (\ref{eq:Eexpand}) and  (\ref{eq:tE}).
For any $\alpha>1$,  $R_A(t)$ satisfies the estimates:
\ba
\sup_{2\Gamma^{-\alpha}\le t\le T} \lan t\ran^{r-1} \ |R_A(t)| 
 &\le& C_1e^{-\Gamma^{-\delta}} \sup_{0\le\tau\le \M}|E(\tau)|\ +\ 
  C\ \Gamma^{-1}\ \sup_{\M\le \tau\le T}
  \left(\lan\tau\ran^{r-1}|E(\tau)|\right) ,
  \label{eq:prop4b}\\
  \sup_{0\le t\le 2\Gamma^{-\alpha}} \lan t\ran^{r-1} \ |R_A(t)|
   &\le&\ D\ \Gamma^{-\alpha r}\ \sup_{0\le \tau\le 2\Gamma^{-\alpha} }\ |E(\tau)| 
   \label{eq:prop4c}
\ea
\end{prop}
\noindent
\underline{Proof.}
To prove (\ref{eq:prop4b}) we begin with    (\ref{eq:Aeqn3})
\be  \D_tA(t)\ =\ \rho(t)A(t) - \Gamma A(t) + E(t), \label{eq:Edef}
\ee
where $E(t)$ is given by (\ref{eq:Eexpand}).    
Let
\be
\tilde{A}(t) \equiv e^{- \int^t_0 \rho(s)ds} A(t).\label{eq:tAdef}
\ee
Then, $\tilde{A}$ satisfies the equation
\begin{eqnarray}
\D_t\tilde{A} &=& -\Gamma \tilde{A} + \tilde{E}(t) \label{eq:tAeqn1}\\
\tilde{E}(t) &=& e^{-\int^t_0 \rho(s)ds} E(t). \label{eq:tE}
\end{eqnarray}
Solving (\ref{eq:tAeqn1}) we get
\ba
\tilde{A}(t)\ &=&\ e^{-\Gamma t} \tilde{A}(0)\ +\
 \int^t_0 e^{-\Gamma(t-s)} \tilde{E}(s)\ ds
\label{eq:tAinteqn}\\
&\equiv&\ e^{-\Gamma t} \tilde{A}(0)\ +\ R_A(t).
\ea

Since the real part of  $\rho(t)$ is of the general form of 
 $c_1 \cos{\omega_1 t} + c_2 \sin{\omega_2 t}$, 
  its integral is bounded uniformly in $t$.  Therefore, 
$$   C^{-1} |\tilde{A}(t)| \le |A(t)| \le C |\tilde{A}(t)| 
 \ {\rm for \ some} \ \ C > 0,$$
and similarly for $E(t), \tilde{E}(t)$.  
 It is therefore sufficient to estimate 
  $\tilde{A}(t)$, in terms of $\tilde{E}(t).$  

  \nit {\bf Remark:} Estimates of 
   $R_a(t)$, which appears in the statement of Theorem 2.2,
   are  related to those for  $R_A(t)$ via :
   \be
   \left|R_a(t)\right|\ =\ \left|e^{-\int_0^t\rho(s)\ ds} \ R_A(t)\right| 
	\ \le\ C\ \left|R_A(t)\right|\nn\ee
\medskip

 Note that on any fixed time interval, $0\le t\le 2M$, one has by
 Gronwall type estimates
  that, $|\tilde{A}(t)|$ and $\|\wmi\phi_d(t)\|$ are
   bounded by a constant 
 depending on the initial data and $M$. This bound is, in general,
  exponentially
 large in $M$.  We shall now focus on obtaining appropriate bounds for
 these quantities on $t-$intervals, $[2M,T)$, which are independent of
 $T$.

 From (\ref{eq:tAinteqn}) we have for any $M>0$: 
\ba
|\tilde{A}(t)|\ &\le& |A(0)|e^{-\Gamma t} + 
 \int^M_0 e^{-\Gamma(t-s)}  |\tilde{E}(s)| ds +
\int^t_M e^{-\Gamma(t-s)} |\tilde{E}(s)|\ ds\no\\
&=&\ |A(0)|e^{-\Gamma t}\ +\ I_1(t)\ +\ I_2(t).
 \label{eq:tAM}
\ea
Set 
 $$M=\Gamma^{-\alpha},\ \ \alpha>1.$$
We now estimate the terms $I_1(t)$ and $I_2(t)$
in (\ref{eq:tAM}) for $2\M\le t\le T$.
\ba
\lan t\ran^{r-1}\ I_1(t)\ &=&\ \lan t\ran^{r-1}\ \int^M_0 e^{-\Gamma(t-s)}
|\tilde{E}(s)| ds\no\\
&\le&\ \lan t\ran^{r-1}e^{-{1\over2}\Gamma t }
 \cdot \int^M_0 e^{-\Gamma ({1\over2}t-s)}\ ds\
\cdot \sup_{0\le\tau\le\M} |\tilde E(\tau)|\no\\
&\le&\ 
\sup_{2\M\le\tau\le T}(\lan t\ran^{r-1}e^{-\Gamma {t\over2} }) \cdot
 C\Gamma^{-1} \cdot 
  \sup_{0\le\tau\le\M} |\tilde E(\tau)|\no\\
 &\le&\ Ce^{-\Gamma^{-\delta}}\ \sup_{0\le\tau\le\M} |\tilde E(\tau)|,
\label{eq:I1est1}
\ea 
for some $\delta>0$. Therefore, 
\be
\sup_{2\M\le\tau\le T}\left(\lan t\ran^{r-1}\ I_1(t)\right)
\ \le \  Ce^{-\Gamma^{-\delta}}\ \sup_{0\le\tau\le\M} |\tilde E(\tau)|
\label{eq:I1est2}
\ee
We estimate $I_2(t)$ on the interval $2\M\le t\le T$ as follows:
\be
\lan t\ran^{r-1}\ I_2(t)\ \le\  
 \lan t\ran^{r-1}\  \int_{\M}^t e^{-\Gamma(t-s)} 
  \lan s\ran^{-r+1} \ ds\   
   \sup_{\M\le\tau\le T}\left(\lan\tau\ran^{r-1}\tilde{E}(\tau)\right)
   \label{eq:I2est1}
\ee
The integral is now bounded above using the estimate
\be
\lan t\ran^{r-1}\ \int_{\M}^t\ e^{-\Gamma(t-s)}\ \lan s\ran^{-r+1}\ ds\ 
    \le\ C\Gamma^{-1},\ t\ge2\M.
\label{eq:integralestimate}
\ee

This gives
\be
\sup_{2\Gamma^{-\alpha}\le t\le T}\lan t\ran^{r-1}\ I_2(t)\ 
 \le\  C\Gamma^{-1}\ 
\sup_{\M\le\tau\le T}\left(\lan\tau\ran^{r-1}\tilde{E}(\tau)\right)
\label{eq:I2est2}
\ee

Assembling the estimates (\ref{eq:I1est2}) and (\ref{eq:I2est2})
yields estimate (\ref{eq:prop4b}) of Proposition 5.1.
 Estimate (\ref{eq:prop4c}) is a simple consequence of 
  the definition of $R_A(t)$.

\section{Dispersive Estimates and Local Decay.}

In this section we prove the local decay of $\phi_d$ and the decay in 
 time of the remainder terms in the ODE of Section 4.  
  To this end we will repeatedly use the following:

\begin{lem} For any $\eta\in [0,r-1]$ 
\be
\left\| \int^t_0 \wmi e^{-iH_0(t-s)} \Pc f(s)ds \right\|\ \le\  
 C\lan t\ran^{-\eta} \sup_{0 \le\tau\le t} 
 \left( \lan\tau\ran^{\eta}\|\wpl f(\tau) \|\right)
 \label{eq:localdecayintegral}
\ee
and
\be
\left\| \int^t_0 \wmi e^{-iH_0(t-s)} \Pc (H_0-\la_0-\mu-i0)^{-1}f(s)\ ds 
 \right\|\  
 \le\ C\lan t\ran^{-\eta} \sup_{0\le\tau\le t}
 \left(\lan\tau\ran^{\eta}\left\|\wpl f(\tau)\right\|\right). 
  \label{eq:localdecaysingular}
\ee
\end{lem}

\noindent
\underline{Proof.}  The proof follows from the assumed local decay estimates on $e^{-iH_0t}$; see {\bf (H3)}. Namely, using that $r>2$, 
\ba
\left\| \int^t_0 \wmi e^{-iH_0(t-s)} \Pc f(s)\ ds \right\|\  
&\le&\  \int_0^t \| \wmi e^{-iH_0(t-s)}\ \Pc \wmi \|_{{\cal L}({\cal H})}
 \lan s\ran^{-\eta}\ ds\no
\\
&&\ \cdot\sup_{0\le\tau\le t} \left(\lan\tau\ran^{\eta} 
	 \|\wpl f(\tau )\| \right)\no 
  \\
  &\le&\ C\int_0^t \lan t-s \ran^{-r+1}\lan s\ran^{-\eta}\ ds\  
\sup_{0 \le\tau\le t}\left( \lan\tau\ran^{\eta} \|\wpl f(\tau )\| \right)\no
 \\
 &\le& C\lan t\ran^{-\eta} \sup_{0 \le\tau\le t}\left(\lan\tau\ran^{\eta}
  \left\| \wpl f(\tau ) \right\|\right)\no 
\ea
which proves (\ref{eq:localdecayintegral}).  
 The proof of (\ref{eq:localdecaysingular}) is identical, and uses the
singular local decay estimate of {\bf (H3)}. 

We now define the norms 
\be
 [A]_\alpha(T)\ =\ \sup_{0\le\tau\le T} 
  \lan\tau\ran^\alpha |A(\tau)| \nonumber
 \ee
 and 
 \be
 [\phi_d]_{LD,\alpha}(T)\ =\ \sup_{0\le\tau\le T} \lan\tau\ran^\alpha 
 \| \wmi\phi_d(\tau) \|\no
\ee
Then we have

\begin{prop}
For any $T>0$,  
\begin{equation}  
[\phi_d]_{LD,r-1}(T) \le C 
  \left(\ \|\wpl\phi_d(0)\|\ +\  
  ||| W |||\  [A]_{r-1}(T)\ \right) .
\label{eq:phidestimate}
\end{equation}
\end{prop}

\noindent
\underline{Proof.}
 From equation (\ref{eq:intphideqn}) we get, using the assumed 
local decay estimate for $e^{-iH_0t}$ and (\ref{eq:localdecayintegral}),
 
\ba
\left\| \wmi\phi_d(t) \right\|\ &\le&\  \sum_{j=0}^2\ \|\wmi\phi_j(t)\|\no\\
 &\le&\ C\lan t\ran^{-r+1} 
 \left\| \wpl \phi_d(0) \right\|
+\  C\lan t\ran^{-r+1}\  [A]_{r-1}(t)\ 
 \sup_{0\le s\le t}\|\wpl W(s)\psi_0\|\no\\
&+&\  C\ ||| W |||\ \lan t\ran^{-r+1}\left[\phi_d\right]_{LD,r-1}(t).
\label{eq:phixyz}
\ea

Since $\|\wpl W(s)\psi_0\|\le ||| W |||\ \|\psi_0\|\ =\ |||W|||$ 
 and $||| W |||$ is 
 assumed to be small, multiplying both sides of this last equation 
  by $\lan t\ran^{r-1}$ and taking supremum over $t\le T$ yields
  (\ref{eq:phidestimate}).
  \ \ \ \ \ []

We now estimate $E(t)$.

\begin{prop} Let $T>0$. For any $\eta\in [0,r-1]$: 
\be
  [E]_{\eta}(T)\  
\le\ C \left(\ |||W|||^2\  |A(0)|\  +\  |||W|||\  \|\wpl\phi_d(0)\|\
+\ |||W|||^3\  [A]_{\eta}(T)\ \right)    
\label{eq:Eestimate}
\ee
\end{prop}

\noindent
\underline{Proof.}

$E(t)$ is defined in (\ref{eq:Edef}) and (\ref{eq:Aeqn3}). From these
equations it is seen that we need to estimate
 the following terms:

\ba
&(\alpha_1)& |R_{1a}| = |A(0)| \ 
 \left|\left(\beta \psi_0, e^{-iH_0t}(H_0 - \la_0 + \mu)^{-1} \Pc  
 \beta \psi_0\right)\right|\no \\
&(\alpha_2)& |R_{1b}| = |A(0)| \left| 
 \int^t_0 \left(\beta\psi_0, e^{-iH_0(t-s)} e^{i(\mu-\la_0)s} 
 (H_0-\la_0+\mu)^{-1} \D_sA(s)\Pc \beta\psi_0 \right)ds \right|\no \\
&(\alpha_3)& |R_3| = {1\over2} |A(0)| \ \left|(\beta \psi_0, 
 e^{-iH_0t}(H_0 - \la_0 - \mu -i0)^{-1} \Pc  \beta \psi_0)\right|\no\\
&(\alpha_4)& |R_4| =  {1\over2} \left| \int^t_0 (\beta \psi_0, e^{-iH_0(t-s)}
 (H_0 - \la_0 - \mu -i0)^{-1} e^{-i(\la_0+\mu)s} \D_sA(s) 
 \Pc\beta \psi_0)\ ds\right|\no\\
&(\alpha_5)& |(\beta\psi_0, \phi_0)| = |(\beta\psi_0, e^{-iH_0t} \phi_d(0))|
 \no\\
&(\alpha_6)& |(\beta\psi_0, \phi_2(t))| = 
 \left| \int^t_0 (\beta\psi_0, e^{-iH_0(t-s)}\cos{\mu s}\ \Pc\ 
 \beta \phi_d(s)\ ds) \right|. \no
\ea
  The estimates of 
 $(\alpha_j)$ repeatedly use Lemma 6.1. Let $\eta\in [0,r-1]$. 

\nit
\un{Estimation of $(\alpha_1)$:}

\ba
|R_{1a}| &\le& |A(0)| 
\left(\wpl\beta\psi_0,\  \wmi e^{-iH_0t}(H_0-\la_0+\mu)^{-1}\Pc\wmi\ 
	\wpl\beta\psi_0\right)\no\\
&\le& C |A(0)|\ |||W|||^2\ \lan t\ran^{-r+1}
\label{eq:alpha1est}
\ea
by local decay estimates and our assumptions on $H_0$, since $\la_0 - \mu$ 
 is not in the spectrum of $H_0$.

\medskip
\noindent
\un{Estimation of $(\alpha_2)$}

From (\ref{eq:Edef}) we have that
\be
|\D_sA(s)| \le C |||W||| \ |A(s)| + |E(s)|  \label{eq:DAofsestimate} 
\ee
since $\rho$ and $\Gamma$ are, respectively, linear and quadratic in $|||W|||$.

Applying Lemma 6.1 to $(\alpha_2)$ we then get
\ba
|R_{1b}|\ &\le&\ 
|A(0)| \left| \int_0^t \left(\wpl\beta\psi_0,\ \wmi e^{-iH_0(t-s)} 
 (H_0-\lambda_0 +\mu)^{-1}\Pc \wmi\ \D_sA(s)\ \wpl\beta\psi_0\right)\
 ds\right|\no\\
&\le&\ C ||| W|||^2\ \lan t\ran^{-\eta} 
 \left( |||W|||\ [A]_{\eta}(t)\ +\ [E]_{\eta}(t)\right).
 \ea

\noindent
\underline{Estimation of  $(\alpha_3)$}
Since $\la_0 + \mu \in \sigma(H_0)$,
 we use (\ref{eq:singularldest})
  together with the procedure used to estimate $(\alpha_2)$.
 The result is:

\be |R_3| \le C |A(0)|\ |||W|||^2 \; \lan t\ran^{-r+1}. 
\label{eq:R3est}
\ee

\medskip
\noindent
\underline{Estimation of  ($\alpha_4$)}

This term is estimated in a manner similar to  $\alpha_2$ and
$\alpha_3$:

\be
|R_4| \le c|||W|||^2  \left\{ |||W|||\ [A]_{\eta}(t)  + 
 [E]_{\eta}(t) \right\} \lan t\ran^{-\eta}.
 \no\ee

\medskip
\noindent
\underline{Estimation of  $(\alpha_5)$:}

Since, by definition, $\phi_\alpha(0) = \Pc\phi_d(0)$ 
 we can apply local decay estimates for $e^{-iH_0 t}$  to get
\be
\left|\left(\beta\psi_0, \phi_0(t)\right)\right|\ \le\  
	 C|||W|||\ \lan t\ran^{-r+1} \; 
 \|\wpl\phi_d(0)\|.   
 \label{eq:alpha5est}
 \ee

\medskip
\noindent
\underline{Estimation of $(\alpha_6)$}

Applying Lemma 6.1 as before we get, for $0\le t\le T$, 
\be
\left|\left(\beta\psi_0, \phi_2(t)\right)\right|\ \le C|||W|||^2\  
 \lan t\ran^{-\eta} \; [\phi_d]_{LD,\eta}(T). \label{eq:temp}
\ee
Using Proposition 6.1 to estimate $[\phi_d]_{LD,\eta}(t)$ in (\ref{eq:temp}),
 we get
\be
\left|\left(\beta\psi_0, \phi_2(t)\right)\right|  
 \le C|||W|||^2\ \lan t\ran^{-\eta} \left\{\ \|\wpl\phi_d(0)\|\ +\ 
 |||W|||\  [A]_{\eta}(t)\ \right\}.  
\no\ee

Finally, combining the above estimates for $(\alpha_j),\ 1\le j\le6$, 
 we can now 
 bound $[E]_{\eta}(T)$ for any $\eta\in [0,r-1]$  as follows: 

\be
 [E]_{\eta}(t) \le C \left\{ 
 |||W|||^2\ |A(0)| + |||W|||\ \|\wpl\phi_d(0)\| + |||W|||^2\  [E]_{\eta}(T)
  + |||W|||^3\ [A]_{\eta}(T)\right\} .
\ee
Since $|||W|||$ assumed to be small, Proposition 6.2  follows. \ \ \ \ []
\medskip

We can now complete the proofs of our main results,
 Theorem 2.1 and Theorem 2.2.  To prove the assertions concerning the
 infinite time behavior, the key is to establish local decay
 of $\phi_d$, in particular, the uniform boundedness of $[\phi_d]_{LD,r-1}(T)$.
 This will follow directly from Proposition 6.1 
 if we prove the uniform boundedness 
 $[A]_{r-1}(T)$, or equivalently $[\tilde{A}]_{r-1}(T)$.

\begin{prop}  There exists a constant $C_*$,
 depending on $\|\phi_0\|, \|w_+\phi_0\|,
 |||W|||, \Gamma$  and $r$,  such that for any  $T>0$
\be
[A]_{r-1}(T) \le C_*
  \no\ee
\end{prop}

\noindent
\underline{Proof.}

We begin with the expansion of $A(t)$ given in Proposition 5.1.
Multiplying (\ref{eq:prop4a}) by $\lan t\ran^{r-1}$, and taking the
supremum over $0\le t\le T$ we have:
\be
[A]_{r-1}(T)\ \le\  
 \left( |A(0)|\ \Gamma^{-r+1}\ +\ 
 \sup_{0\le \tau\le 2\Gamma^{-\alpha}} |R_A(\tau)|\ +\ 
  \sup_{2\Gamma^{-\alpha}\le \tau\le T} |R_A(\tau)|\ \right)
 \label{eq:oops}\ee
The right hand side of (\ref{eq:oops}) is estimated using Proposition
5.1.
\be
[A]_{r-1}(T)\ \le\  
 \left( |A(0)|\ \Gamma^{-r+1}\ +\  
  D\ \Gamma^{-\alpha r}\ [E]_0(2\Gamma^{-\alpha})\  +\ 
  C_1\ e^{-\Gamma^{-\delta}}\ [E]_0(2\M)\ +\
 C\ \Gamma^{-1}[E]_{r-1}(T) \right).
 \ee
 Next, we apply Proposition 6.2 which yields:
 \ba
[A]_{r-1}(T)\ &\le&\  
 \left( |A(0)|\ \Gamma^{-r+1}\ +\  
D\ \Gamma^{-\alpha r}\ [E]_0(2\Gamma^{-\alpha})\  +\ 
 C_1 e^{-\Gamma^{-\delta}}\ [E]_0(2\M)\right)\no\\
 &+&\ C\Gamma^{-1}\left( |A(0)| |||W|||^2\ +\ 
 |||W|||\ \|\wpl\phi_d(0)\|\ +\ |||W|||^3\ [A]_{r-1}(T)\right).
\no\ea

 Note that by Proposition 6.2 and the simple bound: 
 \be
 [A]_{r-1}(T)\ \le
  C\ \lan T\ran^{r-1}\ [A]_0(T)\ \le\ C\ \lan T\ran^{r-1}\ \|\phi_0\|, 
  \nn\ee
  $[E]_0(2\M)$ 
 is  bounded in terms of the initial data and $|||W|||$.

By hypothesis {\bf (H6)}, there is a positive constant $\theta_0$, such that  
\be
{|||W|||^3\over\Gamma}\ \le {|||W|||\over\theta_0}.
\no\ee
Therefore, for $|||W|||$ sufficiently small
\be
[A]_{r-1}(T) \le C_* 
\label{eq:this-est}
\ee
(Alternatively, we can obtain (\ref{eq:this-est}) by imposing that
$|||W|||^3\Gamma^{-1}$ be sufficiently small.) 
 Here, $C_*$  depends on 
$\|\phi_0\|, \|w_+\phi_0\|,
 |||W|||, \Gamma$ and  $r$.

This completes the proof of Proposition 6.3 and therewith the $t\to\infty$
asymptotics asserted in Theorems 2.1 and 2.2

\nit {\bf Remark:} This estimation procedure differs from one used in 
\cite{kn:TDRT1}, where the cases $r-1$ large and $r-1$ small 
are treated differently.

It remains to prove the intermediate time estimate (\ref{eq:Raestimate}).
 The ingredients are contained in (\ref{eq:Eestimate}) and its proof.
 First, by (\ref{eq:prop4aa})
 \be
 R_A(t)\ \le C\ \int_0^t\ e^{-\Gamma(t-\tau)}\ \left| E(\tau)\right|\
 d\tau.\label{eq:RA}
 \ee
 Let $T_0$ denote an arbitrary  fixed positive  number. We estimate
 (\ref{eq:RA}) for $t\in [0, T_0\Gamma^{-1}]$. We bound the exponential in
 the integrand by one (explicit integration would give something of order
  $\Gamma^{-1}$), and bound $|E(\tau)|$ by estimating the expressions 
  $(\alpha_j),\ j=1,...,6$ in the proof of Proposition 6.2. First, the
  estimates of Proposition 6.2 for $\alpha_1, \alpha_3$ and $\alpha_5$ are
  useful as is. Integration of the bounds (\ref{eq:alpha1est}),
  (\ref{eq:R3est}) and (\ref{eq:alpha5est}) gives:
  \be
  \int_0^t\ e^{-\Gamma (t-\tau)}\ (\alpha_j)\ \le\ 
   C\ |||W|||\ \|w_+\phi(0)\|,\ j=1,3,5.
   \label{eq:est135}
   \ee
   To estimate the contributions of $(\alpha_2)$ and $(\alpha_4)$, first
   observe that by (\ref{eq:DAofsestimate}) and Proposition 6.2 with
   $\eta=0$ 
   \be
   |\D_sA(s)|\ \le\ C\ |||W|||\ \|w_+\phi(0)\| 
   \nn\ee
   Therefore, using local decay estimates we have:
   \ba
   \int_0^t\ e^{-\Gamma (t-\tau)}\ (\alpha_j)\ d\tau&\le&\ 
	C\ T_0\Gamma^{-1}\ |||W|||^3\  \|w_+\phi(0)\|
	\nn\\
	&\le&\ |||W|||\ \|w_+\phi(0)\|,\ j=2,4.
	\nn\ea
Finally, we come to the contribution of $(\alpha_6)$. We rewrite 
 $(\alpha_6)$ as follows.
\ba
(\alpha_6)\ &=&\  
 \left| \int^t_0 ( \beta\ e^{iH_0(t-s)}\Pc \beta\psi_0, \cos{\mu s}\
  \phi_d(s)\ ds) \right|\nn\\
&=&\  \left| \int^t_0 ( w_+\beta w_+\cdot w_- e^{iH_0(t-s)}\Pc w_-\cdot
  w_+\beta\psi_0, \cos{\mu s}\
   w_-\phi_d(s)\ ds) \right|.\label{eq:alpha6rewrite}\ea
Recall that by (\ref{eq:intphideqn}) 
 $\phi_d\ =\ \phi_0\ +\ \phi_1\ +\ \phi_2$,  
  where $\phi_0(t)=e^{-iH_0t}\phi_d(0)$. Using local decay estimates {\bf
  (H3)}, the
  contribution of the term $\phi_0(t)$ to $(\alpha_6)$ can be bounded by 
  $C\ |||W|||^2\ \|w_+\phi_d(0)\|\ \lan \tau\ran^{-r+1}$. Multiplication of this
  bound by $e^{-\Gamma(t-\tau)}$ and integration with respect to $t$
  gives the bound $C\ |||W|||^2 \|w_+\phi_d(0)\|$.
   To assess the contributions from $\phi_1+\phi_2$,
  note that local decay estimates {\bf (H3)} imply  
  \be 
  \|w_-(\phi_1+\phi_2)\|\ \le\ C\ |||W|||\ \| w_+\phi_0\|.
  \nn\ee
  Putting together the contributions from $\phi_0$ and from
  $\phi_1+\phi_2$, we have:
  \be
  \int_0^t\ e^{-\Gamma (t-\tau)}\ (\alpha_6)\ d\tau
  \ \le\ C\left(\ |||W|||^2\ \|w_+\phi_d(0)\|\ +\ 
   \Gamma^{-1}\ |||W|||^3\ \right)
   \nn\ee
The above estimates and {\bf (H6)} imply 
  (\ref{eq:Raestimate}). This concludes the proof
of Theorem 2.2.

\section{Appendix A:\ General approach to local decay estimates}

\bigskip
Hypothesis {\bf (H3)} for our main theorem is one requiring that our
unperturbed operator, $H_0$, satisfy a suitable local decay estimates.
It is remarked following this hypothesis that, in practice, the
verification of (\ref{eq:singularldest}) using  
 (\ref{eq:localdecay1}). In this section we give an outline to a
very general approach to obtaining estimates of this type,  
 based on a technique
originating in the work of Mourre \cite{kn:Mr}; see also \cite{kn:PSS}.
 In the following general
discussion  
we shall  let   
 $H$ denote self-adjoint operator on a Hilbert space, ${\cal H}$,
keeping in mind that our application is to the unperturbed operator $H_0$.
 Let $E\in\sigma(H)$, and assume that an operator $A$ can be found such
that $A$ is self-adjoint and ${\cal D}(A)\cap {\cal H}$ is dense in
${\cal H}$.
Let $\Delta$ denote an open interval with compact closure.
We shall use the notation:
\be
{\rm ad}_A^n(H)\ =\ [\cdots[H,A],A],\cdots A],\nn
\ee
for the n-fold commutator.

 Assume the two conditions:  
  
  \nit {\bf (M1)}\  The operators  
\be g_\Delta(H)\ [H,A],\ [H,A]\ g_\Delta(H),\ {\rm and}\ 
g_\Delta(H)\ {\rm ad}_A^n(H)\ g_\Delta(H),\ \ 2\le n\le N\ee
can all be extended to a bounded operator on ${\cal H}$. 

\bigskip

\nit {\bf (M2)}\ {\it Mourre estimate}:
\be
g_\Delta (H)\ i[H,A]\ g_\Delta (H)\ \ge \theta\ g_\Delta(H)^2\ +\ K
\ee
for some $\theta >0$ and compact operator, $K$.

\nit{\bf Remark:} The theorems below were orginally proved under more
restrictive conditions
 than {\bf (M1)} \cite{kn:CFKS}, \cite{kn:GS}, \cite{kn:Mr},
 \cite{kn:PSS}, \cite{kn:SS}, \cite{kn:Skibsted}. 
 The more general results stated here can be proved using the 
  approach to velocity bounds in
\cite{kn:HSS}. 
\bigskip

\begin{theo} (Mourre;\  see \cite{kn:CFKS} Theorem 4.9, \cite{kn:BFSS} Lemma
5.4)

\nit Assume conditions {\bf (M1)-(M2)}, with N=2. Then, in the interval
$\Delta$, $H$ can only have 
absolutely continuous spectrum with finitely many eigenvalues
of finite multiplicity. Moreover, the operator
\be 
\lan A\ran^{-1}\ g_\Delta(H)\ (H-z)^{-1}\ \lan A\ran^{-1} 
\ee
is uniformly bounded in $z$, as an operator on ${\cal H}$.
If $K=0$, then there are no eigenvalues in the interval $\Delta$.
\end{theo}

\bigskip
\bigskip

\begin{theo} (Sigal-Soffer;\ see \cite{kn:SS},\cite{kn:GS},
\cite{kn:Skibsted}, \cite{kn:HS}, \cite{kn:HSS})

\nit Assume conditions {\bf (M1)-(M2)} with  $N\ge2$ and 
 $K=0$. Then, for all $\varepsilon >0$
\be
||\ F\left({A\over t}<\theta\right)\ e^{-iHt}\ g_\Delta (H)\psi\ ||_2
\ \le\ C \lan t\ran^{-N + 1 + \varepsilon}\ \|\ \lan A\ran^{N-1}\psi\ \|_2,
\ee
and therefore
\be
||\ \lan A\ran^{-\sigma}\ e^{-iHt}\ g_\Delta (H)\psi\ ||_2\ 
  \le\ C\ \langle t\rangle^{-\sigma}\ \|\ \lan A\ran^{N-1}\psi\ \|_2,
\ee
for $\sigma < N-1$. Here, $F$ is a smoothed out characteristic
function, and  $F\left({ A\over t}<\theta\right)$ is defined by
the spectral theorem.
\end{theo}

\nit{\bf Remark:} $\sigma$ is required to be smaller than $N\over2$ in  
\cite{kn:GS}, \cite{kn:SS}  and smaller than that in \cite{kn:Skibsted}. 
 The above bound $\sigma<N-1$ is
proved in \cite{kn:HSS}; see Theorem 1.1 and the remark below it.

 Let $\Delta_1$ denote an open interval
 containing the closure  of $\Delta$.
\medskip

\begin{cor}
 Assume that $\lan x\ran^{-\sigma}\ g_{\Delta_1}(H)\lan
 A\ran^\sigma$ is  
bounded. Then, in the above theorems we can replace the weight 
 $\lan A\ran^{-\sigma}$ by $\lan x\ran^{-\sigma}$.
\end{cor}
\bigskip

The strategy for using the above results to prove local decay estimates like
that in {\bf (H4)} is as follows. 
 Then 
\ba
\|\ \lan x\ran^{-\sigma}\ e^{-iHt}g_\Delta(H)\psi\ \|_2
\ &=& \|\  \lan x\ran^{-\sigma}g_{\Delta_1}(H) e^{-iHt}g_\Delta(H)\psi\|_2
\nn\\
\ &=& \|\  \lan x\ran^{-\sigma}g_{\Delta_1}(H) \lan A\ran^\sigma\cdot  
\lan A\ran^{-\sigma}
\ e^{-iHt}g_\Delta(H)\psi\ \|\nn\\
&\le&\ \|\  \lan x\ran^{-\sigma}g_{\Delta_1}(H) \lan A\ran^\sigma \| \cdot
 \|\ \lan A\ran^{-\sigma}
 \ e^{-iHt}g_\Delta(H)\psi\ \|\nn\\
 &\le&\ C\ \|\ \lan A\ran^{-\sigma}
  \ e^{-iHt}g_\Delta(H)\psi\ \|\nn\\
 &\le&\ C_1 \|\ F\left({A\over t} <\theta\right)\ \lan A\ran^{-\sigma}
  \ e^{-iHt}g_\Delta(H)\psi\ \|_2\ \nn\\
  &&\  +\ C_2\|\ F\left({A\over t}\ge \theta\ \right) \lan A\ran^{-\sigma}
   \ e^{-iHt}g_\Delta(H)\psi\ \|_2
\label{eq:split}
\ea
Theorem 7.2 is used to obtain the decay of the first term on the right
hand side of (\ref{eq:split}), while we can replace $A$ by $\theta t$
in the second term.


\end{document}